\documentclass{emulateapj} 

\usepackage{graphicx}
\usepackage{color}
\usepackage{float}
\usepackage{subfigure}
\usepackage{longtable}
\usepackage{amsmath}
\usepackage{amssymb}
\usepackage{enumitem}
\usepackage[flushleft]{threeparttable}
\usepackage{hyperref}

\def\thesgr{SGR J1550$-$5418}
\def\thesgrs{SGR J1550$-$5418~}

\begin{document} 

\title{Burst Tails from \thesgrs Observed with Rossi X$-$ray Timing Explorer}
 
\author{Sinem \c{S}a\c{s}maz Mu\c{s}, Ersin G\"o\u{g}\"u\c{s}, Yuki Kaneko, Manoneeta Chakraborty, Berk Ayd$\i$n}

\affil{Sabanc\i~University, Faculty of Engineering and Natural 
  Sciences, Orhanl\i ~Tuzla 34956 Istanbul Turkey}
  
\email{sinemsmus@sabanciuniv.edu}  

\begin{abstract} 
 
We present the results of our extensive search using the Bayesian block 
method for long tails following short bursts from a magnetar, \thesgr,
over all {\it RXTE} observations of the source. We identified four 
bursts with extended tails, most of which occurred during its 2009 burst 
active episode. The durations of tails range between $\sim$13 s and over 3 ks, 
which are much longer than the typical duration of bursts. We performed 
detailed spectral and temporal analysis of the burst tails. We find that 
the spectra of three tails show a thermal nature with a trend of cooling 
throughout the tail. We compare the results of our investigations with the 
properties of four other extended tails detected from SGR 1900+14 and 
SGR 1806$-$20 and suggest a scenario for the origin of the tail in the 
framework of the magnetar model.  
\end{abstract} 
  

\section{Introduction}

\label{sec:intro}

Magnetars $-$ neutron stars powered by their extremely strong magnetic 
fields \citep{duncan92,thompson95} $-$ are distinguished by the emission 
of energetic bursts observed in the hard X-ray/soft gamma-ray band. 
Currently there are 28 sources classified as magnetars (see the magnetar 
catalog\footnote{\url{http://www.physics.mcgill.ca/~pulsar/magnetar/main.html}}, 
\citep{olausenkaspi14} for detailed information). Magnetar bursts can be 
classified according to durations and energetics: short bursts last a 
fraction of a second and involve an isotropic energy of $\lesssim10^{40}$ erg. 
Intermediate events are slightly longer, typically a few seconds, and the emitted 
energy is about 2 orders of magnitude larger. Magnetars emit giant 
flares but very rarely; only three such flares have been observed to 
date. The giant flares are at the extreme of the burst energy scale 
($\gtrsim$10$^{44}$ erg) and relatively long, lasting a few hundreds 
of seconds, during which there are remarkable spectral and temporal 
variations. For a comprehensive list of studies on magnetar bursts, see 
the magnetar burst library\footnote{\url{https://staff.fnwi.uva.nl/a.l.watts/magnetar/mb.html}}.

\thesgr, also known as 1E 1547.0-5408 or PSR J1550-5418, is a magnetar 
with currently the shortest spin period, 2.072 s \citep{enoto10}. The 
spin period and spin-down rate were measured first in the radio band 
\citep{camilo07}. It was first proposed as a magnetar candidate by 
\cite{gelfand07} based on its magnetar-like X-ray spectrum and 
association with a supernova remnant. Identification of its spin 
period and spin-down rate, which implies a magnetic field strength 
of 2.2$\times$10$^{14}$ G, further supported the suggested magnetar 
hypothesis \citep{camilo07}. Although there were implications that 
it has gone through X-ray brightening episodes 
\citep{gelfand07,halpern08}, magnetar-like bursts from the source 
were not observed until 2008 October \citep{krimm08}. \thesgrs 
exhibited other intense bursting episodes in 2009 January and March 
\citep{connaughton09,vonkienlin09}. 

There have been numerous extensive investigations in order to 
understand the burst and persistent X-ray emission properties of 
\thesgr. \cite{israel10}, using {\it Swift} observations of the 2008 October 
burst activation, found that the 2$-$10 keV flux was elevated by 
$\sim$50 times above its quiescent level, and that its pulsed fraction 
has also increased significantly. Spectral analysis of the bursts 
observed with the Burst Alert Telescope (BAT) on board {\it Swift} revealed 
that a blackbody model with temperature of 11 keV represents the 
burst spectra very well \citep{israel10}. \cite{vonkienlin12} 
analyzed the bursts observed with the {\it Fermi} Gamma-ray Burst Monitor 
(GBM) in 2008 October and 2009 March--April. They reported that the 
spectral characteristics of the bursts observed in two active 
episodes separated by about 5 months are different: the 2008 October 
burst spectra are best described with a single-blackbody function, 
while bursts observed in 2009 March--April are better fit with an 
optically thin thermal bremsstrahlung model. They interpreted this 
variation as a reflection of the changes in magnetic field structure 
of the source due possibly to another extreme-intense bursting 
episode that occurred in between these two periods (in January 2009). 
\cite{vanderhorst12} analyzed {\it Fermi} GBM observations of 286 bursts 
detected during a week following 2009 January 22, the most burst-active 
episode of the source. They reported that burst spectra 
can be described equally well with a Comptonized model or 
double-blackbody model. \cite{lin12} additionally used simultaneous 
{\it Swift} observations to analyze the bursts of the same period including 
the soft X-ray band and found that the double-blackbody model represents 
the spectra better than the Comptonized model. 

Besides typical short bursts, there have been reports on more 
energetic events from \thesgr, mostly during its 2009 January active 
phase. \cite{mereghetti09} reported that some energetic bursts 
(with energies as high as 10$^{43}$~erg) detected with {\it INTEGRAL} 
are followed by long emission episodes which are modulated 
with the spin period of the neutron star. \cite{kaneko10} identified 
a 150 s long enhanced persistent emission phase during which 
pulsed signals were detected up to $\sim$110 keV. \cite{kuiper12} 
identified two events in the Rossi X-ray Timing Explorer ({\it RXTE}) 
observations of the same active episode and called them 
`mini outbursts' due to their long emission ($\sim$450 s and 
$\sim$130 s) at a much lower intensity than the bursts, but clearly 
above the persistent emission level. Here, we call the latter events 
bursts with extended tails.

SGR 1900+14 and SGR 1806$-$20 have shown energetic bursts 
with extended tails; extended tails typically last a few hundreds of seconds
but can be as high as thousands of seconds. In all of these cases, 
the spectral properties of the tail emission are different than from
those of the bursts, as well as those of the persistent emission; the 
tail spectra are well fitted with a blackbody model with decreasing 
temperature throughout the course of the tail, which implies a cooling 
thermal component on the surface, possibly heated by the initiating 
burst \citep{ibrahim01,lenters03,gogus11}. \cite{lenters03} and 
\cite{gogus11} showed that the total energy contained in the extended 
tails accounts for a constant percentage of the initiating burst event; 
it is $\sim$2$\%$ for SGR 1900+14 \citep{lenters03} while for the two 
detected tails in SGR 1806$-$20 the ratios are $\sim$0.34$\%$ and 
$\sim$0.63$\%$ \citep{gogus11}.

Motivated by the detection of extended tails from other magnetars, 
and having already identified two bursts with tails from \thesgrs 
\citep{kuiper12}, we extensively searched for extended tails following 
bursts in all available {\it RXTE} observations of \thesgr. However, 
detection of these tails on short timescales is not optimal due to 
variations of the background emission and, sometimes, the existence of 
hundreds of bursts in the active episode. Thus, in the work presented 
here, we applied a Bayesian block algorithm \citep{scargle13}, which can 
detect local variabilities more robustly to search for burst tails from 
\thesgr. In the following, the {\it RXTE} observation details are found in 
Section \ref{sec:data}, and in Section \ref{sec:bayesian} we explain 
our search methodology for the detection of tails, as well as our search results. 
We present the results of our detailed spectral and temporal investigations 
of the identified burst tails in Sections \ref{sec:spectral} and 
\ref{sec:temporal}, respectively. We discuss the physical implications of 
our results and compare with the properties of extended burst tails 
observed from other magnetars in Section \ref{sec:discuss}.


\begin{figure*}
\begin{center}
\includegraphics[scale=0.7]{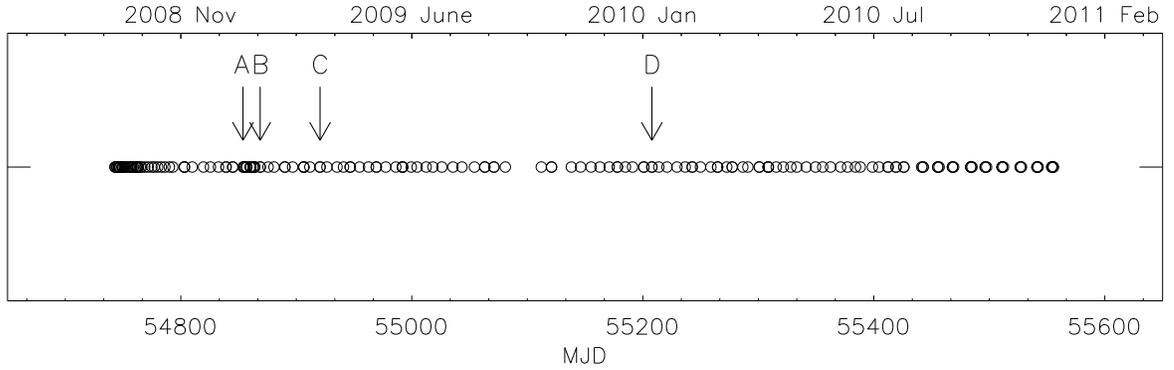}
\caption{Distribution of \thesgrs observations performed by {\it RXTE}. 
Downward arrows indicate the tails detected with {\it RXTE} in events 
A, B, C, and D. Exposures of these observations range from 0.5 to 
17 ks, with a mean of about 3.7 ks.}
\end{center}
\label{fig:obs}
\end{figure*}


\section{Observations and Data Analysis}
\label{sec:data}

{\it RXTE}, which was operational from 1995 December to 2012 January,
had two instruments on board: Proportional Counter Array (PCA), 
detecting photons in the energy range 2$-$60 keV, and The High 
Energy X-ray Timing Experiment (HEXTE) covering an energy range 
of 15$-$250 keV. PCA had five proportional counter units (PCU)
labeled from 0 to 4, each consisting of one propane veto, three 
xenon, and one xenon veto layer. In the 2$-$10 keV energy range, 
PCU sensitivity limit was 4$\times$$10^{-12}$ erg s$^{-1}$ 
cm$^{-2}$, and telemetry rate as high as 20,000 counts s$^{-1}$ 
\citep{jahoda06}. 

Here we used all available archival data obtained biweekly between
2008 October and 2010 December (191 observations, total exposure of 
$\sim$702 ks; see Figure \ref{fig:obs} for the time distribution 
of these observations). In our investigations, we used data collected 
with the PCA only. For timing analysis, we converted the arrival 
times to the time at the Solar System barycenter using the source 
coordinates of R.A. = 15$^{h}$50$^{m}$54$^{s}$.11 and 
decl. = $-$54$^{\circ}18'23''.7$ given by \cite{camilo07}.


\begin{table}
\caption{Details of Observations That Contain Extended Tails.}
\begin{threeparttable}
\centering
\begin{tabular}{l c c c}
\hline\hline
Event	& Date& Time (UTC)\tnote{a}  & Active PCUs \\
\hline
A 	& 2009 Jan 22 & 22:48:45.44   & 2\\
B 	& 2009 Feb 06 & 18:29:03.15   & 2, 3\\
C 	& 2009 Mar 30 & 14:13:06.20   & 1, 2\\
D 	& 2010 Jan 11 & 21:12:23.40   & 2\\ [1ex]
\hline
\end{tabular}
\label{table:details}
  \begin{tablenotes}
        \item[a] Denotes the start of the event.\\
    \end{tablenotes}
\end{threeparttable}
\end{table}



\begin{figure*}
\subfigure{\includegraphics[scale=0.55]{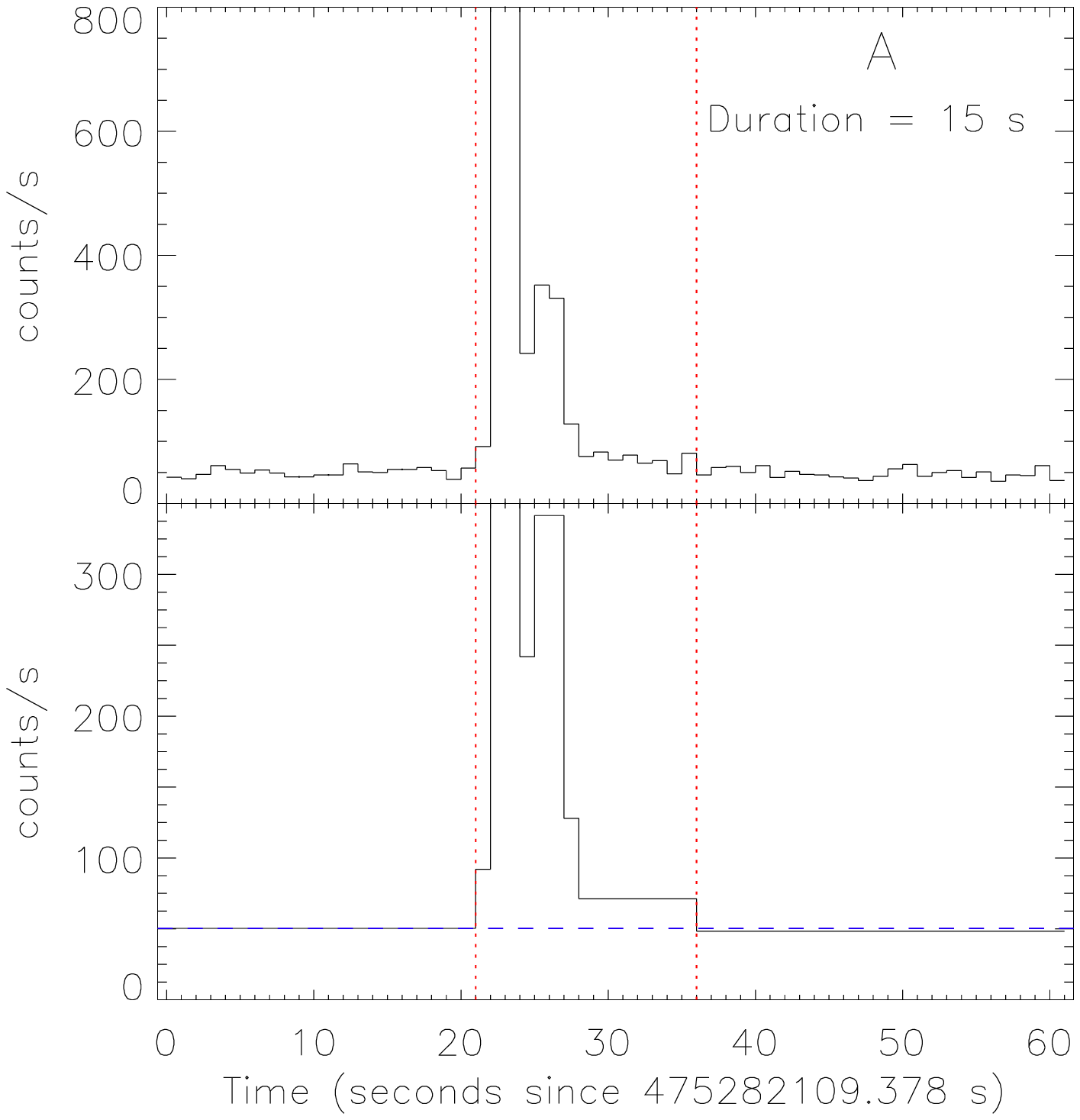}}\quad
\subfigure{\includegraphics[scale=0.55]{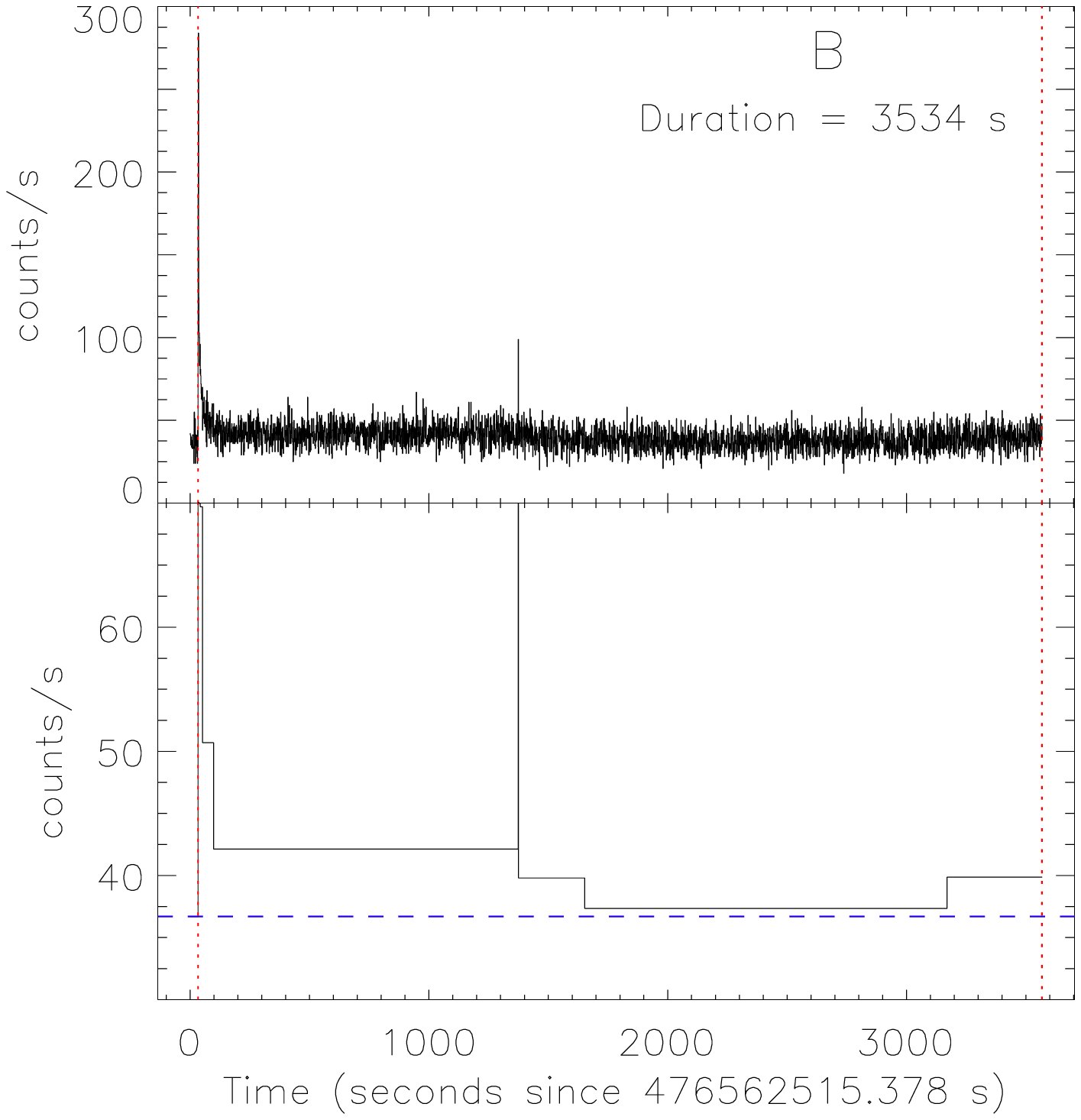}}\quad
\subfigure{\includegraphics[scale=0.55]{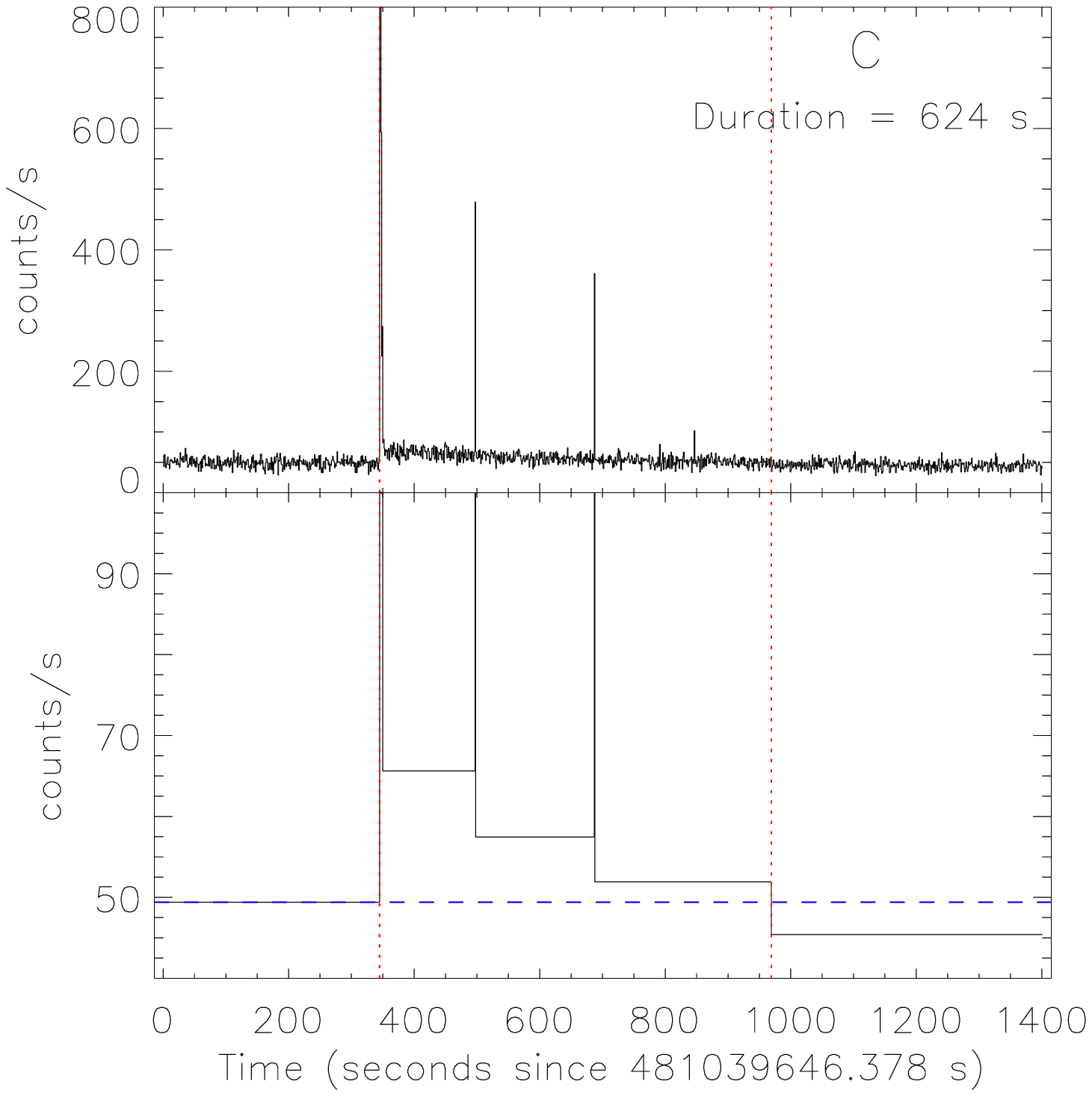}}\quad
\subfigure{\includegraphics[scale=0.55]{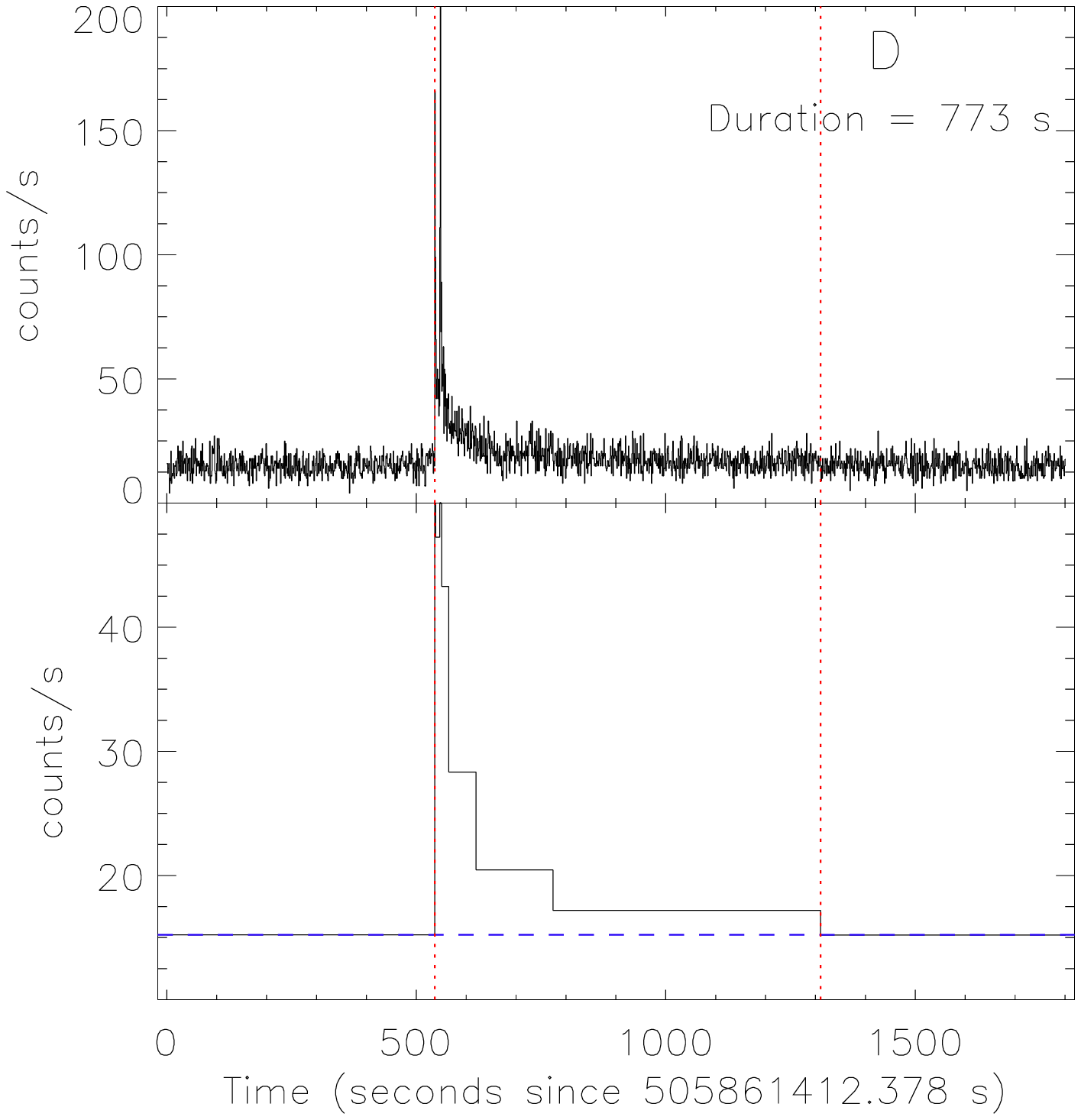}}
\caption{Top panel: 2$-$20 keV light curves of the four events 
with 1 s resolution. The four events are identified here with 
labels A$-$D and indicated in each panel. Bottom panel: Bayesian 
block representation of the light curve. The blue dashed line 
represents the background level. Red vertical dotted lines 
indicate the start and end times of the events. For event B 
the tail end time is the end of the data (see text).}
\label{fig:bayesian}
\end{figure*}


\section{Search for Extended Tails}
\label{sec:bayesian}

We implemented a Bayesian blocks based algorithm to identify 
extended burst tails, which immediately follow the bursts with count rate 
much lower than that of the burst, but still higher than the 
count rate of the pre-burst data. The Bayesian block algorithm is a 
segmentation method in order to detect local variabilities in 
time series data by separating the data into blocks of statistically 
significant variations that maximize the likelihood value \citep{scargle13}.
This method was used to search for weak 
magnetar bursts by \cite{lin13} and resulted in successful 
detection of the dimmest bursts observed from magnetars.

To optimize the extensive search, we took a two-step approach: 
a signal-to-noise ratio (S/N) burst identification followed by a 
Bayesian block tail search, which we describe below. First, we 
searched for burst candidates in the 0.125 s binned light curves 
of the source using an S/N criterion with significance 5.5$\sigma$ 
above background in the 2$-$20 keV energy range. The S/N search 
identified 878 burst candidates in total. Then, for each of the 
burst candidates we regenerated the light curves with 1-s resolution, 
for the interval between 200 s before and 1000 s after the burst 
time. We then applied the Bayesian block algorithm to the 1 s 
light curves and obtained the Bayesian block representation of the 
light curves. Using those, we searched for the increase in the count 
rate that matches the burst times found by the S/N search. 
Once found, the beginning of the identified block was taken as 
the event start time, and the count rate of the preceding block was 
assigned as the background level. Then, the first decrease in the 
count rate after the event start time below the assigned background 
level determines the putative end of the tail. If the algorithm 
cannot determine the end of the tail within the selected data 
segment (from T$-$200 to T+1000 s), we extend the post-burst interval 
by 200 s and perform the search again. For our analysis we run two 
iterations, and if the algorithm still cannot assign an end to the 
tail, we conclude that there is no tail associated with that particular 
burst. We note that the assigned background level can be affected by 
the existence of data gaps, and in such cases the tail may have been 
missed in the search. Therefore, to account for such cases, we also
artificially elevated the background level to 30$\%$ of
the burst candidate's count rate and performed the search again.
The tail end time in such a case was the end of the observation.
We eliminated the tail candidates that lasted less than
1 s since extended tails of our interest have much longer durations.
Finally, we excluded false detections that are due to known data 
anomalies. 

Based on these criteria, we identified a total of four bursts with 
extended tails from \thesgr, which we labeled A$-$D in chronological 
order for convenience in this paper. In Table \ref{table:details}, 
we provide observational details of these events. Note that our sample 
includes the two events (events C and D) visually identified and mentioned in 
\cite{kuiper12}. In Figure \ref{fig:bayesian}, we present the 
light curves of the bursts with extended tails, as well as their Bayesian block 
representations. Among these events, the shortest event duration 
determined by the algorithm is 15 s, and longest duration is 3534 s.
For the longest event, the tail end time corresponds to the 
end of the observation (see Figure \ref{fig:bayesian} top right panel).


\begin{table}
\caption{Duration and count rates of bursts and tails in 2$-$25 keV band.} 
\begin{threeparttable}
\centering
\begin{tabular}{l c c }
\hline\hline
Event   &  Duration (s) & Counts s$^{-1}$\tnote{a}   \\
\hline

{\bf A}    &	     &		      \\
Burst	   & 1.83    & 5573 (5524)    \\
	
Tail 	   & 17.00   & 102 (53)   \\			  
\hline
{\bf B}    &	     &                \\
Burst	   & 0.12    & 767 (737)  \\
	
Tail 1	   & 47.07   & 68 (28)   \\	  
	
Tail 2	   & 146.91  & 47 (7)    \\
\hline
{\bf C}    &	     &		      \\
Burst 1	   & 0.35    & 18450 (18400)   \\

Burst 2	   & 0.17    & 2206 (2147)    \\
		  
Burst 3	   & 0.20    & 1270 (1218)     \\
	
Tail 1	   & 148.22  & 74 (19)         \\
	
Tail 2	   & 187.83  & 63 (8)          \\
					  
Tail 3	   & 279.80  & 57 (2)         \\					  
\hline
{\bf D}	   &	     &	              \\
Burst 1	   & 0.30    & 203 (190)  \\

Burst 2	   & 0.45    & 340 (329)  \\

Tail 1	   & 113.30  & 31 (14)    \\
			  
Tail 2	   & 647.40  & 20 (3)     \\  								     
\hline
\end{tabular}
  \begin{tablenotes}
        \item[a] The values in parentheses are background-subtracted count rates.\\
    \end{tablenotes}
\end{threeparttable}
\label{tab:specdet}
\end{table}


\section{Spectral Analysis of Tails and Associated Bursts}
\label{sec:spectral}

We investigated spectral properties of the extended tails 
and the bursts that are associated with these tails in order to 
uncover their emission properties and compare them with the 
detected tails from SGR 1900+14 and SGR 1806$-$20. We used HEASOFT
v6.16 to perform data extraction. First, we applied standard
filtering to the PCA data (Earth occultations, South Atlantic Anomaly 
passages, electron contamination, etc.). We performed spectral
fits using XSPEC software version 12.8.2 \citep{arnaud96} in
energy range 2$-$25 keV, and we used an interstellar H column density 
of 3.4 $\times$ 10$^{22}$ cm$^{-2}$ which was obtained from 
{\it Swift} data analysis of the source \citep{lin13}. 

We obtained the background spectra for the bursts from pre-burst 
data and, when possible, also from post-burst data. For the tails 
we extracted the background spectra only from the pre-burst 
data. In all cases we used all layers of operating PCUs and 
finally grouped the burst and tail spectra such that each 
spectral bin would contain at least 20 counts (except the burst 
in event B and burst 1 in event D, which have less counts than other
bursts; these are grouped to contain at least 10 counts).

The durations of the four events with tails A, B, C, and D are 
15, 3534, 624 and 773 s, respectively, determined based on 
Bayesian block representation. As mentioned 
in the previous section, the reason for the long tail duration 
of event B is that the algorithm could not find a block that goes 
below the assigned background level before the end of {\it RXTE} orbit. 
We note, however, that the emission after $\sim$200 s is quite weak; 
in fact, inclusion of data after $\sim$200 s in the spectral analysis
did not significantly alter the parameters. Therefore, we limit 
our investigation by that time.

\begin{figure}
	\subfigure{\includegraphics[scale=0.5]{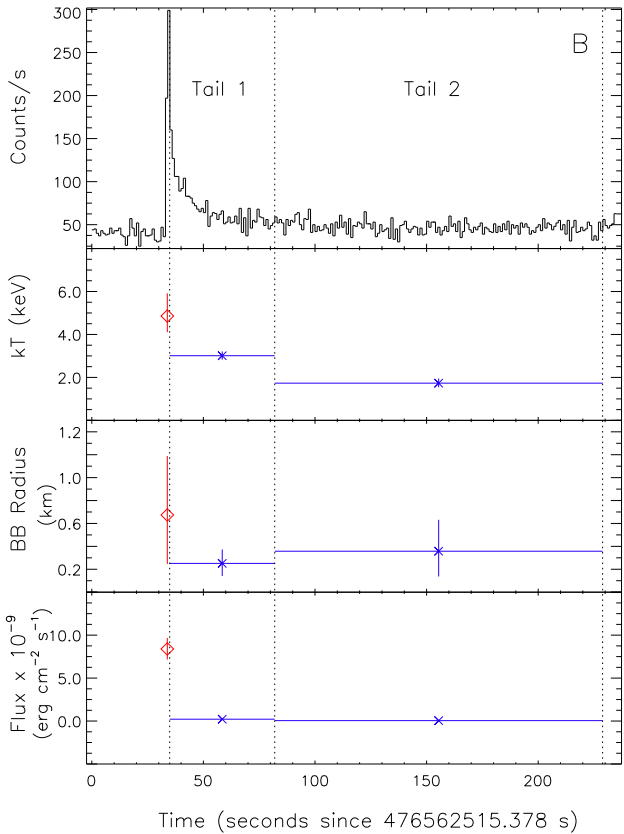}}
	\caption{Spectral parameters of the burst and tail of event B. 
	Top panel: 1 s binned light curve of the event. Dotted vertical 
	lines indicate the time intervals of tail segments. Second panel: 
	blackbody temperature of the burst (red diamond) and tail segments 
	(blue crosses). Third panel: blackbody radii of the burst and tail segments. 
	Bottom panel: unabsorbed fluxes of the burst and tail segments.}
	\label{fig:eventb}
\end{figure}

For each of the four events, we analyzed the burst and tail spectra 
separately. In the cases of events B, C, and D, the sufficiently long 
durations of the tails enabled us to investigate the spectral 
evolutions throughout the tails. To this end, we divided the 
tails of B and D into two segments, and tail of event C into three 
segments, separated by the other two bursts in this tail (see Figures 
\ref{fig:eventb}, \ref{fig:eventc} and \ref{fig:eventd}), and performed 
time-resolved spectral analysis. We present the durations and count 
rates of the burst and tail spectra in Table \ref{tab:specdet}.

For modeling the burst and tail spectra, we employed a set of spectral 
models that are commonly used for magnetar spectral analysis: single 
blackbody, double blackbody, optically thin thermal bremsstrahlung, 
power law, cutoff power law, as well as combinations of these models 
(see, e.g., \cite{israel10,vonkienlin12,vanderhorst12,lin12}).  
Below, we describe the spectral analysis results of each event.  
We also present the best-fit spectral parameters determined
by the $\chi^{2}$ statistics in Table \ref{tab:specparams} and 
spectral evolution of events B, C, and D in Figures 
\ref{fig:eventb}, \ref{fig:eventc} and \ref{fig:eventd}.
We note that uncertainties are reported at the 1$\sigma$ level 
throughout the paper.

\subsection{Results of Spectral Analysis}

\subsubsection{Event A}
The burst in the beginning of event A was saturated due to 
high number of burst photons. We therefore excluded the time 
intervals (a total of $\sim$0.15 s) during which the count 
rate exceeded 18,000 c s$^{-1}$PCU$^{-1}$. The burst spectrum 
is described best with the cutoff power-law model with a photon index 
of 0.68$\pm${0.08} and cutoff energy 14.24$^{+1.96}_{-1.57}$ keV 
($\chi^{2}_{red}$ = 1.14 with 49 degrees of freedom (dof)). 
We found that the tail of this event is fitted with a power-law 
model the best: index of 1.37$^{+0.10}_{-0.11}$ 
($\chi^{2}_{red}$ = 0.84 with 34 dof). The 2$-$25 keV unabsorbed 
fluxes of the burst and the tail are $>$9.87$\pm$0.12$\times10^{-8}$ 
and 9.23$\pm0.44\times10^{-10}$~erg~cm$^{-2}$~s$^{-1}$, respectively. 
The corresponding energies, assuming isotropic emission and the source 
distance of 5 kpc \citep{tiengo10} are $>$5.39$\times$10$^{38}$~erg 
and 4.69$\times$10$^{37}$~erg for the burst and the tail, 
respectively. Note that the burst energy is a lower bound since 
the detector was saturated, and its true energy content was higher 
than our estimate. 

\subsubsection{Event B}
The burst in event B can be fitted with a single-blackbody model 
with a temperature of 4.86$^{+1.06}_{-0.75}$ keV 
($\chi^{2}_{red}$ = 0.43 with 6 dof). The time-integrated spectrum 
of the tail of this event is also best fitted with a blackbody model 
of kT = 2.22$^{+0.22}_{-0.19}$ keV ($\chi^{2}_{red}$ = 0.98 with 53 
dof). Note that the power-law model for this time-integrated tail yields 
a similar fit statistics: $\chi^{2}_{red}$ of 1.01 with an index of 
1.51$\pm$0.18. We also analyzed the spectra of each segment of the 
tail and found that both segments are described well with a blackbody 
model of changing temperature of 3.01$^{+0.20}_{-0.18}$ and 
1.73$^{+0.22}_{-0.19}$ keV (see Figure \ref{fig:eventb}). The 
corresponding radii of the blackbody emitting region are 
0.25$^{+0.12}_{-0.11}$ km and 0.36$^{+0.27}_{-0.22}$ km,
respectively (given d = 5 kpc). 

\subsubsection{Event C}
\begin{figure}
	\subfigure{\includegraphics[scale=0.5]{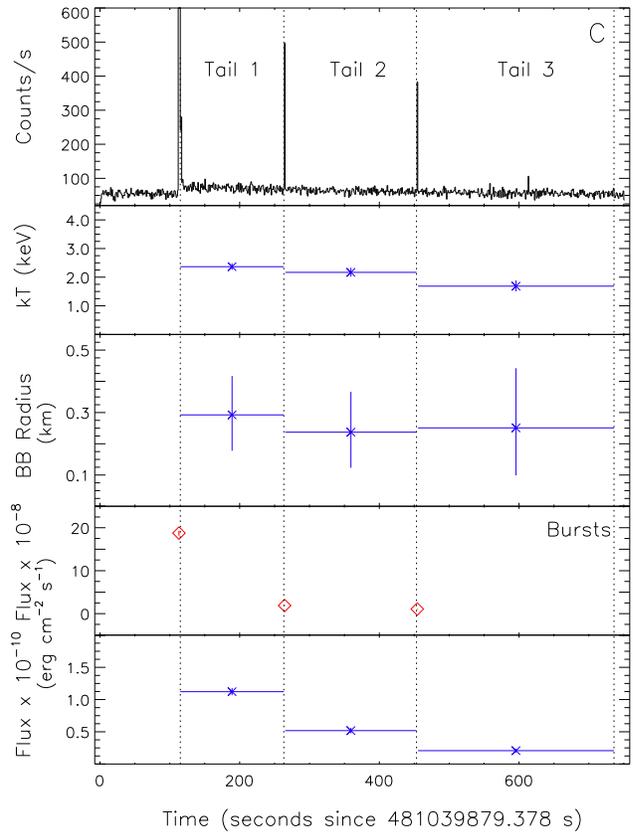}}
	\caption{Spectral parameters of the bursts and tail of event C. 
	Dotted vertical lines indicate the time intervals of tail 
	segments. Top panel: 1 s binned light curve of the event. 
	Second panel: blackbody temperature of the tail segments. 
	Third panel: blackbody radii of the tail segments. Fourth 
	panel: unabsorbed fluxes of the bursts. Bottom panel: 
	unabsorbed fluxes of the tail segments.}
	\label{fig:eventc}
\end{figure}

Event C contains three bursts, one in the beginning of the 
event, the second one $\sim$150 s after the first burst, and the last 
one separated by $\sim$190 s from the second burst (see Figure 
\ref{fig:eventc}). The first burst has also saturated the detector due 
to a high number of incoming burst photons, thus, we excluded the 
saturated portion of the burst (a total of $\sim$0.18 s), applying 
the same count rate criterion used for the event A. For this burst, 
a combination of blackbody and power-law models provided the best fit 
($\chi^{2}_{red}$ = 1.24 with 52 dof); the blackbody temperature is 
15.46$^{+4.37}_{-2.58}$~keV and power-law index is 2.01$\pm0.14$. 
This corresponds to a blackbody radius of 0.64$^{+0.45}_{-0.40}$~km. 
The slightly large $\chi^{2}_{red}$ obtained in this best fit was 
actually due to the data around 14 keV. Excluding the energy range of 
11$-$16 keV from the spectral analysis improves the fit statistics. 
We will investigate this burst in detail in \c{S}a\c{s}maz Mu\c{s}, S. 
et al. (2015, in preparation). Second and third bursts can be fitted well 
with a power-law model with photon indices 1.43$\pm$0.11 
($\chi^{2}_{red}$ = 1.01 with 13 dof) and 1.79$\pm0.16$
($\chi^{2}_{red}$ = 1.00 with 8 dof), respectively. As for the tail, 
a blackbody model fits the time-integrated tail spectrum very well 
($\chi^{2}_{red}$ = 1.18 with 54 dof). The blackbody temperature is 
2.12$^{+0.11}_{-0.10}$ keV, and corresponding radius is 
0.24$^{+0.11}_{-0.10}$ km. The three segments of this tail can also 
be fitted with the blackbody model, resulting in temperatures of 
2.37$\pm0.11$, 2.17$^{+0.16}_{-0.15}$ and 
1.69$^{+0.20}_{-0.18}$~keV, respectively (see Figure \ref{fig:eventc}). 
These correspond to blackbody emitting region radii of 
0.29$^{+0.12}_{-0.11}$, 0.24$^{+0.13}_{-0.11}$ and 
0.25$^{+0.19}_{-0.15}$~km.

\subsubsection{Event D}

\begin{figure}
	\subfigure{\includegraphics[scale=0.5]{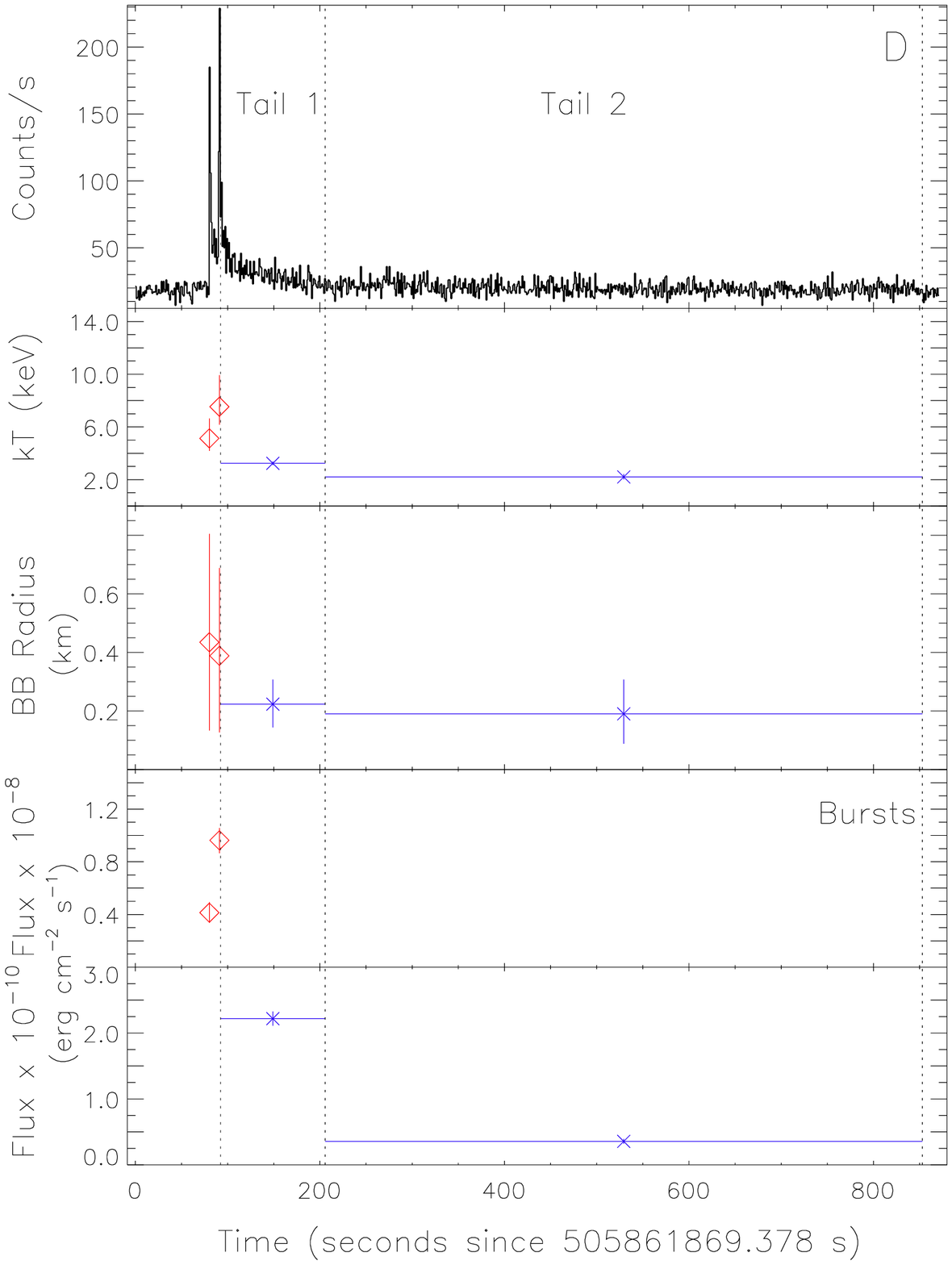}}
	\caption{Spectral parameters of the bursts and tail of event D. 
	Dotted vertical lines indicate the time intervals of tail 
	segments. Top panel: 1 s binned light curve of the event. 
	Second panel: blackbody temperature of the bursts (red diamonds) 
	and tail segments (blue crosses). Third panel: blackbody radii 
	of the bursts and tail segments. Fourth panel: unabsorbed 
	fluxes of the bursts. Bottom panel: unabsorbed fluxes of tail 
	segments.}
	\label{fig:eventd}
\end{figure}

At the onset of event D, there are two bursts separated by 
$\sim$11 s. The first burst is best fitted with a single blackbody 
with a temperature of 5.13$^{+1.50}_{-0.94}$ keV but still 
resulted in large $\chi^{2}$ ($\chi^{2}_{red}$ = 1.56 with 4 dof). 
Further investigation of this burst revealed a possible spectral 
feature around 14 keV. Similar spectral features around the same 
energy have been observed in other magnetars 
\citep{gavriil02,woods05,gavriil06,an14}. Inclusion of a Gaussian 
line improves the fit statistics significantly, but energetics are 
comparable with or without the Gaussian line. This will also be 
investigated in detail in \c{S}a\c{s}maz Mu\c{s}, S. et al. (2015, in preparation).   
The second burst of this event is fitted with a single blackbody 
with kT = 7.53$^{+2.39}_{-1.36}$ keV ($\chi^{2}_{red}$ = 0.34 with 
5 dof). The time-integrated tail spectrum is also fitted well with 
a blackbody model with temperature 2.59$^{+0.13}_{-0.12}$~keV 
($\chi^{2}_{red}$ = 0.91 with 53 dof). The corresponding blackbody 
radius is calculated as 0.18$^{+0.08}_{-0.07}$~km. Similar to the 
other events, we found that the tail can be modeled with a blackbody  
of decreasing temperature, 3.23$^{+0.14}_{-0.13}$~keV in the first and 
2.20$^{+0.21}_{-0.18}$ keV in the second segment 
(see Figure \ref{fig:eventd}). The corresponding blackbody radii are 
0.22$\pm0.08$ and 0.19$^{+0.12}_{-0.10}$~km, respectively.


\begin{table*}
\caption{Best-fit Spectral Parameters (2$-$25 keV) for the Bursts and the Tails}
\begin{threeparttable}
\centering
\setlength{\tabcolsep}{2.5pt}
\begin{tabular}{l c c c c c c c c c}
\hline\hline
Event	  	 &Model$^{a}$&   kT    	&   Index		    & BB Radius 	    & Flux  (erg~cm$^{-2}$~s$^{-1}$)	     & \multicolumn{2}{c}{Isotropic Energy (erg)}	& $\chi^{2}_{red}$(dof)$^{c}$  \\
        	 && (keV)   		&			    & (km)		    	& (2$-$25 keV) Unabsorbed		&(2$-$10 keV) & (2$-$25 keV)			     &  					     \\ 
\hline

{\bf A}  && & & & & & \\

Burst		&Cutoff PL$^{b}$& $-$		& 0.68$\pm$0.08     & $-$		    	    & 9.87$\pm0.12\times10^{-8}$   	      & 2.43$\pm0.03\times10^{38}$& 5.40$\pm0.07\times10^{38}$&1.14(49)  \\	       
\\         		         		
Tail		&PL& $-$			& 1.37$^{+0.10}_{-0.11}$    & $-$		    & 9.23$\pm0.44\times10^{-10}$  	      & 2.11$\pm0.16\times10^{37}$& 4.69$^{+0.23}_{-0.22}\times10^{37}$& 0.84(34)  \\

\hline
{\bf B} && & & & & &  \\

Burst		&BB& 4.86$^{+1.06}_{-0.75}$& $-$			    & 0.67$^{+0.52}_{-0.43}$ & 8.39$^{+1.28}_{-1.23}\times10^{-9}$     & 7.42$^{+1.25}_{-1.23}\times10^{35}$&3.01$^{+0.46}_{-0.44}\times10^{36}$  & 0.43(6)  \\ 	 
\\         		         		
Tail 1+2	&BB& 2.22$^{+0.22}_{-0.19}$& $-$			    & 0.28$^{+0.22}_{-0.18}$ & 8.06$^{+0.78}_{-0.75}\times10^{-11}$    & 3.17$\pm0.31\times10^{37}$	   & 4.68$^{+0.46}_{-0.44}\times10^{37}$ & 0.98(53)  \\
\\
Tail 1		&BB& 3.01$^{+0.20}_{-0.18}$ & $-$			    & 0.25$^{+0.12}_{-0.11}$ & 2.14$^{+0.15}_{-0.14}\times10^{-10}$    & $-$				    & $-$				  & 0.91(98)  \\
\\              		
Tail 2		&BB& 1.73$^{+0.22}_{-0.19}$& $-$			    & 0.36$^{+0.27}_{-0.22}$ & 4.71$^{+0.65}_{-0.64}\times10^{-11}$    & $-$				    & $-$				  & 0.91(98)  \\

\hline
{\bf C} && & & & & &  \\

Burst 1		&BB+PL& 15.46$^{+4.37}_{-2.58}$& 2.01$\pm$0.14   	    & 0.64$^{+0.45}_{-0.40}$ & 1.88$^{+0.06}_{-0.05}\times10^{-7}$  	& 6.12$\pm0.16\times10^{37}$	    & 1.97$^{+0.07}_{-0.06}\times10^{38}$ & 1.24(52) \\		 
\\
Burst 2		&PL& $-$			& 1.43$\pm$0.11 	    & $-$		     & 1.89$\pm0.11\times10^{-8}$  		& 4.50$\pm0.33\times10^{36}$	    & 9.61$^{+0.55}_{-0.54}\times10^{36}$& 1.01(13)   \\
\\
Burst 3		&PL& $-$			& 1.79$\pm$0.16    	    & $-$		     & 1.08$\pm0.07\times10^{-8}$  		& 3.70$^{+0.39}_{-0.37}\times10^{36}$& 6.45$\pm0.43\times10^{36}$& 1.00(8)   \\
\\
Tail 1+2+3	&BB& 2.12$^{+0.11}_{-0.10}$& $-$			    & 0.24$^{+0.11}_{-0.10}$ & 5.06$^{+0.28}_{-0.27}\times10^{-11}$ 	& 6.60$\pm0.35\times10^{37}$&
9.32$^{+0.51}_{-0.50}\times10^{37}$ 	& 1.18(54)  \\
\\
Tail 1		&BB& 2.37$\pm$0.11         & $-$			    & 0.29$^{+0.12}_{-0.11}$ & 1.12$^{+0.06}_{-0.05}\times10^{-10}$     & $-$				    & $-$				& 1.18(162)  \\
\\
Tail 2		&BB& 2.17$^{+0.16}_{-0.15}$& $-$			    & 0.24$^{+0.13}_{-0.11}$ & 5.18$^{+0.42}_{-0.41}\times10^{-11}$ 	& $-$				     & $-$				  & 1.18(162)  \\
\\
Tail 3		&BB& 1.69$^{+0.20}_{-0.18}$& $-$			    & 0.25$^{+0.19}_{-0.15}$ & 2.07$^{+0.30}_{-0.29}\times10^{-11}$ 	& $-$				     & $-$				 & 1.18(162)  \\

\hline
{\bf D} && & & & & & \\

Burst 1		&BB& 5.13$^{+1.50}_{-0.94}$& $-$			    & 0.43$^{+0.37}_{-0.30}$ & 4.14 $^{+0.80}_{-0.77}\times10^{-9}$ 	& 8.53$\pm1.80\times10^{35}$	     & 3.71$^{+0.72}_{-0.69}\times10^{36}$ & 1.56(4)  \\
\\
Burst 2		&BB& 7.53$^{+2.39}_{-1.36}$ & $-$			    & 0.39$^{+0.30}_{-0.26}$ & 9.63$^{+0.94}_{-0.98}\times10^{-9}$  	& 1.96$^{+0.36}_{-0.34}\times10^{36}$& 1.30$\pm0.13\times10^{37}$  & 0.34(5)  \\
\\
Tail 1+2	&BB& 2.59$^{+0.13}_{-0.12}$& $-$			    & 0.18$^{+0.08}_{-0.07}$ & 6.03$^{+0.32}_{-0.31}\times10^{-11}$ 	& 7.83$\pm0.42\times10^{37}$ & 	     1.37$\pm0.07\times10^{38}$  & 0.91(53)  \\
\\
Tail 1		&BB& 3.23$^{+0.14}_{-0.13}$& $-$			    & 0.22$\pm$0.08 	     & 2.22$\pm0.11\times10^{-10}$ 		& $-$				      & $-$				     & 1.12(103)  \\
\\
Tail 2		&BB& 2.20$^{+0.21}_{-0.18}$& $-$			    & 0.19$^{+0.12}_{-0.10}$ & 3.56$^{+0.36}_{-0.35}\times10^{-11}$     & $-$				     & $-$				    & 1.12(103)  \\
\hline
\end{tabular}
  \begin{tablenotes}
        \item[a] PL: power-law model; BB: blackbody model.
	\item[b] Cutoff energy of this model is 14.24$^{+1.96}_{-1.57}$~keV.
	\item[c] We present the simultaneous fit results for the tail segments.
    \end{tablenotes}
\end{threeparttable}
\label{tab:specparams}
\end{table*}



\section{Temporal Analysis}
\label{sec:temporal}

\subsection{Phases of Bursts}

\begin{figure}
	\subfigure{\includegraphics[scale=0.5]{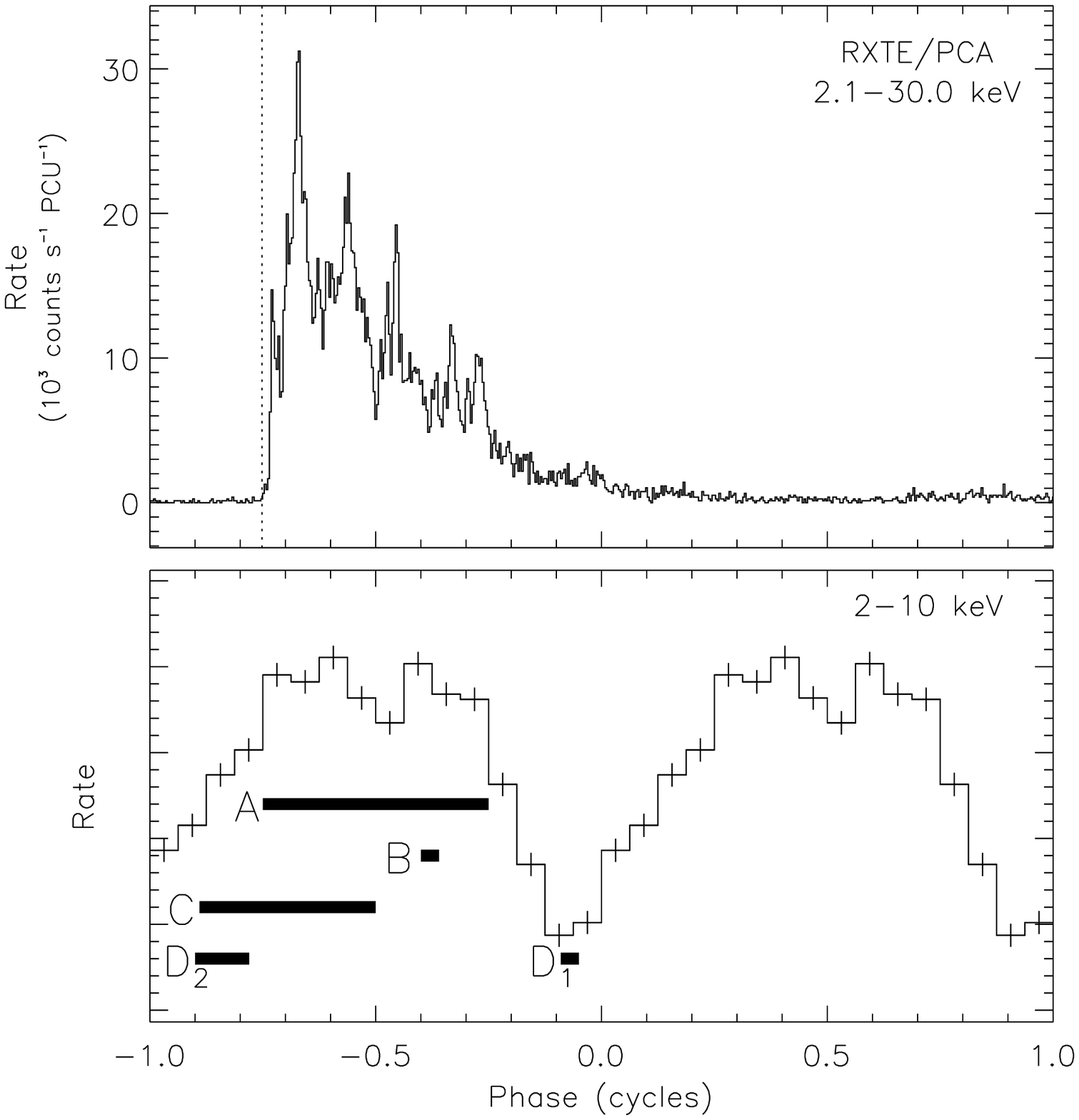}}
	\caption{Top panel: light curve and pulse profile of the tail 
	detected in event A. Black bars in the pulse profile indicate the 
	phase interval of the other events. D$_{1}$ and D$_{2}$ denote the 
	two bursts observed in event D.}
	\label{fig:phase}
\end{figure}

To investigate whether bursts that exhibit extended tails have any 
dependence on the spin phase of the neutron star, we have calculated 
the corresponding phases of the four particular bursts studied here. 
For this purpose, we used contemporaneous  phase-connected spin 
ephemerides of \cite{dib12} for the events A and B, and spin 
ephemerides or frequencies provided by \cite{kuiper12} for the 
remaining two events. We find that the energetic burst of event A 
starts at the spin phase, $\phi$ of 0.25, which is the beginning of 
the peak plateau of the pulse profile (see Figure \ref{fig:phase}), 
continues throughout the pulse peak, and declines rapidly after 
$\phi$ of 0.75, which is the end of the peak plateau. Event B 
spans the spin phase interval of 0.60 to 0.64. Event C starts at 
$\phi$ = 0.11, which is the rising part of the pulse profile, and ends 
somewhere during the peak of the persistent emission. 
Finally, event D has two spikes separated by 11 s: the first one 
spans between $\phi$ of 0.91 and 0.95, which is within the minimum 
phase of the pulse profile, and the second one spans between 
$\phi$ = 0.10 and 0.22, which is again during the rising portion of 
the pulse profile.


\begin{figure}
	\centering
	\subfigure{\includegraphics[scale=0.5]{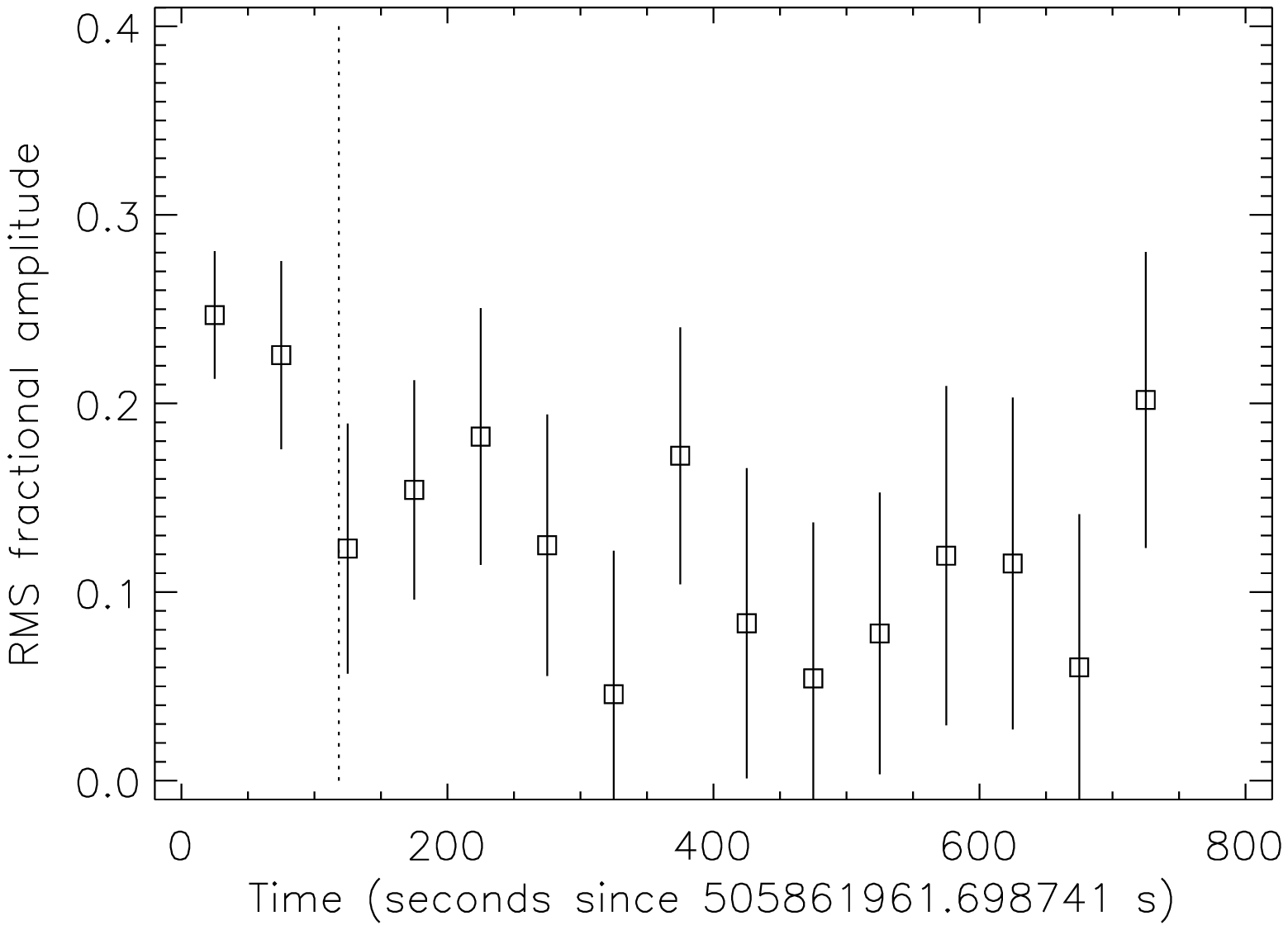}}
	\caption{Variation of rms fractional amplitude during 
	the tail of event D. The rms fractional amplitudes were 
	calculated at each 50 s time bin during the tail. The 
	dotted vertical line represents the separation between the 
	tail segments considered for the spectral analysis.}
	\label{fig:amp_eventd}
\end{figure}


\subsection{Dependence of Pulse Amplitude on Burst Ignition}

Energetic bursts preceding extended tails may induce observable 
changes in the pulse properties of the neutron star. In order to 
investigate this for \thesgr, we compared the pulse amplitudes 
prior to the onset of each of the four events and of those during 
the tails following the bursts. We created the phase-folded light 
curves in the 2$-$20 keV energy range considering only PCU 2 for the 
entire tail and the pre-burst emission of each of the four bursts using 
the spin ephemerides provided by \cite{dib12} and \cite{kuiper12} 
accordingly. We modeled the folded profiles using a sinusoidal 
model with a harmonic and computed the rms fractional amplitude 
from the profiles. For event A, the fractional rms amplitude 
before the burst and during the tail were 1.58$\pm$0.6$\%$ and 
17.66$\pm$2.73$\%$, respectively. For event B the pre-burst emission 
available in the same {\it RXTE} orbit was very short, and we estimated the 
pre-burst pulsation using the data taken during the previous orbit. 
For this burst, we find that the pulsation amplitudes during pre-burst 
and tail were 2.32$\pm$1.15$\%$ and 13.68$\pm$2.38$\%$, respectively. 
For event C, we obtained the pulse amplitude before the burst as 
6.51$\pm$2.50$\%$ and in the tail it became 5.07$\pm$1.39$\%$. 
For the final event D, the amplitudes in the pre-burst emission and 
the tail were 8.16$\pm$2.25$\%$ and 12.20$\pm$1.75$\%$, respectively. 

It was shown for SGR 1900+14 that the pulse fraction increases 
immediately after the burst and varies with time in the tail of the 
burst \citep{lenters03}. In order to check whether the rms amplitude 
of pulsation varies during the tail itself for \thesgr, we considered 
smaller time segments during the tail of all the bursts except the 
first one and computed the phase-folded profile and rms fractional 
amplitudes in each of the short time segments. For event D, the rms 
amplitude is significantly high ($\sim$$0.25$) at the very beginning 
of the tail just following the burst and then quickly decreases 
(see Figure \ref{fig:amp_eventd}). The high rms amplitude in the first 
50 s of the tail complies with the significant detection of high-power 
pulsation by \cite{kuiper12} in the first 52 s since the onset of 
that particular burst. After the first 100 s, the fractional rms amplitude 
value and its variation are consistent with those observed for the 
pre-burst emission. For event B the average rms amplitude showed a 
marginal increase during the early episodes of the tail compared to its 
pre-burst emission, though the increase was not as significant as 
in the case of event D. For event C, the rms fractional amplitude of 
pulsation did not vary significantly, remaining generally consistent 
with the variation of pulse amplitude during the pre-burst emission. 
To give a measure of the variation of the amplitude of pulsation, 
we quote the mean and standard deviation of the pulse amplitudes 
calculated over the finer time segments. The average fractional rms amplitudes 
during pre-burst and the entire duration of tails for the 
three events are: 6.24$\pm$4.04$\%$ and 16.68$\pm$1.26$\%$ for event B, 
7.93$\pm$4.03$\%$ and 6.90$\pm$3.39$\%$ for event C, and 9.64$\pm$2.80$\%$ 
and 13.25$\pm$6.32$\%$ for event D. Note that the errors are calculated 
from the deviations of amplitudes over the short time segments, over 
which the pulsed signal and fractional rms amplitude could only 
be poorly constrained.

\subsection{Search for QPOs in the Tails of Bursts}

We investigated the power spectra during the tails of the four events 
to search for quasi-periodic oscillations (QPOs) that have been 
first observed from SGR 1806$-$20 and SGR 1900+14 among magnetars
\citep{israel05,strohmayer05,watts06}.
For this purpose, we computed the Leahy-normalized 
power spectra over each 1 s segment during the tail in the energy ranges 
2$-$10 and 10$-$30~keV. For QPO detection, we chose a threshold of 
3$\sigma$ significance level, considering the number of trials as the 
number of frequency bins and the number of power spectra searched. Using 
our criterion, we did not detect any significant QPO features in the 
individual or averaged power spectra during the burst tails, and the 
power spectra were consistent with Poisson noise. To further quantify 
the temporal behavior of the burst tails, we calculated the total noise 
rms power over the energy bands 2$-$10 and 10$-$30 keV. We divided the 
burst tails in 15 temporal segments and averaged over them to get the 
resultant  power spectra in each band. We obtained the total rms noise 
in the 10$-$1020 Hz interval for the two energy bands for event A as 
51.09$\%$ and 51.05$\%$, respectively; for event B as 53.28$\%$ and 
53.11$\%$, respectively; for event C as 54.24$\%$ and 53.24$\%$, 
respectively; and finally for event D as 51.96$\%$ and 51.46$\%$, 
respectively. We also computed the confidence level and corresponding 
chance occurrence probability of the highest power detected in the power 
spectrum of each tail, again in the two energy bands. As we present in 
Table \ref{tab:power}, there is no indication of periodic or 
quasi-periodic processes in the tail of these events.


\begin{table*}
\caption{Significance levels of the maximum power observed in the power
spectra in the 10$-$1020 Hz frequency range.}
\centering
\begin{tabular}{c|ccccc}
\hline
Event 		
		& Total rms & Chance &Number of & Confidence & Significance  \\
		&Noise Power & Probability & Trials &  Level ($\%$) & ($\sigma$)  \\
\hline
		&	&	&	Energy Range 2$-$10 keV\\
		
\hline
A & 51.09 & 1.49$\times10^{-3}$ & 1144 & 18.19 & 0.23  \\ 
 
B & 53.28 & 1.01$\times10^{-5}$ & 13061 & 87.65 & 1.54  \\ 

C & 54.24 & 9.44$\times10^{-5}$ & 38924 & 2.54 & 0.03  \\ 
 
D & 51.96 & 7.64$\times10^{-7}$  & 51220 & 96.16 & 2.07  \\
\hline
  &	&  & Energy Range 10$-$30 keV \\
\hline
A & 51.05 & 3.58$\times10^{-4}$ & 1144 & 66.38 & 0.96 \\

B & 53.11 & 5.74$\times10^{-5}$ & 13061 & 47.26 & 0.63 \\

C & 53.24 & 2.91$\times10^{-5}$ & 38924 & 32.21 & 0.41 \\

D & 51.46 & 1.98$\times10^{-5}$  & 51220 & 36.27 & 0.47 \\
\hline
\end{tabular}  
\label{tab:power}
\end{table*}


\section{Discussion}
\label{sec:discuss}

We have searched for extended burst tails using the Bayesian blocks 
technique in a large collection of {\it RXTE} data of \thesgr. We 
identified four events with durations of 15, 3534, 624, and 
773 s, and we have studied spectral and temporal properties of 
these four events. We compare below general properties of these 
events with one another, as well as those of extended burst tails 
detected from other magnetars, namely, SGR 1900+14 and SGR 1806$-$20.

The tail of event A has a much shorter duration ($\sim$13 s). 
Moreover, unlike events B, C and D, its X-ray spectrum has 
non-thermal character; it is described the best with a power-law 
model of index 1.37, resembling the SGR burst spectral shape 
in the {\it RXTE}/PCA passband. The other three burst tails have thermal 
spectra, described with a blackbody model, at temperatures 
around 2$-$3 keV, and more importantly showing 
a clear trend of declining blackbody temperature over the course of 
the tail. According to the magnetar model, the repeated bursts from 
SGR sources are likely due to fracturing of the solid neutron star 
crust by magnetic stress \citep{thompson95} or magnetic reconnection 
\citep{lyutikov03}. In either case, energetic electron$-$positron pair 
plasma would be induced into the magnetosphere which would be 
observed as energetic bursts. Note the important fact that event A 
was detected on 2009 January 22, that is the day \thesgrs was most 
burst active, and many energetic and relatively long bursts were also 
detected (see \cite{mereghetti09}, \cite{savchenko10}). Therefore, 
event A is probably an exceptionally long burst, and the identified 
tail is most likely the continued emission of this prolonged event.

Extended burst tails were also seen from other magnetars: two from 
SGR 1900+14, soon after its giant flare in 1998 August 
and intermediate flare in 2001 April \citep{ibrahim01,lenters03}, 
and two from SGR 1806$-$20 during the 2003$-$2004 burst-active episode 
prior to its giant flare in 2004 December  \citep{gogus11}. Energies 
of the bursts leading to extended tails in SGR 1900+14 were at the highest 
of the scale for typical magnetar bursts. On the other hand, preceding 
bursts of SGR 1806$-$20 tails were not the highest-energy ones: there 
were a lot of other energetic bursts that were not followed by tails. 
Similarly, we found that the energy of the tail triggering burst in \thesgrs 
could be as low as 3.01$\times10^{36}$~erg, and there were a lot 
of more energetic bursts \citep{mereghetti09,vanderhorst12}. 
The extended tail phenomenon is, therefore, a special case in magnetar 
bursts; there should be a minimum energy injection to ignite a tail, 
but not all bursts that are more energetic than the threshold would 
lead to extended tails. Besides, unlike the other magnetars, in the 
case of \thesgrs the ratios of total energy contained in the tails
and in the bursts seem to vary from event to event. Hence, the 
question of what ignites extended tails cannot be answered simply 
with the energetics of the main burst, as already pointed out by 
\cite{gogus11}.

An important common spectral property of all extended tails observed 
now from three sources is that they all exhibit thermal spectra and 
the temperatures decline, that is, cooling throughout the tail. The 
spectral cooling behavior in \thesgrs is not as significant as in the 
cases of SGR 1900+14 and SGR 1806$-$20 tails, but still evident. The 
corresponding blackbody emitting area of \thesgrs tails remains fairly 
constant around 0.2$-$0.3 km for events B, C, and D. Similar to the 
earlier tails detected, \thesgrs extended tails are also exhibiting 
the cooling of a heated portion of the neutron star crust. Given the 
energetics argument above and the spectral nature of tails, we suggest 
that the source of heating is the bombardment of the neutron star 
surface with returning pairs in the trapped fireball \citep{thompson95}, 
which could not efficiently radiate away. Therefore, the consequential 
cause of the generation of extended tails would depend on how efficiently 
the trapped plasma in the magnetosphere radiates away.

As far as the temporal investigations of the extended tails are 
concerned, we do not find any phase dependence of leading bursts, 
and also any evidence of QPOs in the high-frequency domain, which 
are observed from the giant flares of SGR 1900+14 and SGR 1806$-$20, 
and are attributed to torsional oscillations of the neutron star 
crust \citep{israel05,strohmayer05,watts06}. Apart from other causes, 
the QPO phenomenon likely requires a much higher burst energetic 
threshold than that of extended tails to excite torsional modes. 
Interestingly, we find a significant pulsed amplitude increase in the 
very early portion of event D, which has one of the least burst energetics. 
This burst energy argument conditionally strengthens our hypothesis of 
surface heating with returning pairs because as more energy is 
imparted back to the system, the energy content of radiated portion 
would naturally be less.

\section*{Acknowledgments}
We thank the referee for helpful comments and suggestions. 
This project is funded by  the Scientific and Technological Research 
Council of Turkey (T\"UB\.ITAK grant 113R031).

\label{lastpage} 
\end{document}